\documentclass[aps,prd,12pt,preprintnumbers,groupedaddress,floatfix,nofootinbib]{revtex4}

\usepackage{amsmath}
\usepackage{graphicx}
\usepackage{color}

\def\beq{\begin{eqnarray}}
\def\eeq{\end{eqnarray}}
\def\bea{\begin{eqnarray}}
\def\eea{\end{eqnarray}}

\begin{document}

\preprint{DCP-07-02}

\title{{A Model of Neutrino and Higgs Physics at the Electroweak Scale}}
\author{Alfredo Aranda}
\email{fefo@ucol.mx}
\affiliation{Facultad de Ciencias, Universidad de Colima,\\
Bernal D\'{i}az del Castillo 340, Colima, Colima, M\'exico}
\affiliation{Dual C-P Institute of High Energy Physics}

\author{Omar Blanno}
\email{pos00625@alumnos.fcfm.buap.mx}
\affiliation{Facultad de Ciencias Fisico-Matem\'aticas, BUAP \\
Apdo. Postal 1364, C.P.72000 Puebla, Pue, M\'exico}

\author{J. Lorenzo D\'iaz-Cruz}
\email{lorenzo.diaz@fcfm.buap.mx}
\affiliation{Facultad de Ciencias, Universidad de Colima,\\
Bernal D\'{i}az del Castillo 340, Colima, Colima, M\'exico}
\affiliation{Facultad de Ciencias Fisico-Matem\'aticas, BUAP \\
Apdo. Postal 1364, C.P.72000 Puebla, Pue, M\'exico}
\affiliation{Dual C-P Institute of High Energy Physics}

\begin{abstract}
    \noindent
    We present and explore the Higgs physics of a model that in
    addition to the Standard Model fields
    includes a lepton number violating singlet scalar field. Based on the
    fact that the only experimental data we have so far for physics beyond
    the Standard Model is that of neutrino physics, we impose a
    constraint for any addition not to introduce new higher scales.
    As such, we introduce right-handed neutrinos
    with an Electroweak Scale mass. We study the Higgs decay $H \rightarrow \nu
    \nu$ and show that it leads to different signatures compared to those in the Standard
    Model, making it possible to detect them and to probe the nature of their couplings.
\end{abstract}

\maketitle

\newpage

\section{Introduction}
Neutrino physics has received a tremendous amount of experimental
input in the last
decade~\cite{neutrinoresults1,neutrinoresults2,neutrinoresults3,
neutrinoresults4,neutrinoresults5,neutrinoresults6,neutrinoresults7}.
Neutrino oscillations are now completely determined and thus
neutrinos are massive. On the theoretical side, the origin of
neutrino masses and their observed patterns (for the neutrino mass
squared differences) as well as the mixing angles still represent a
mystery~\cite{vallerev}. There are some ideas that have been widely
used in order to explore the situation, like the Zee
model~\cite{zee} or the see-saw mechanism~\cite{seesaw1,seesaw2} in
its several incarnations~\cite{seesaw3}, but we are far from a
profound understanding. Most of the actual realizations of these
mechanisms postpone much of the desired knowledge to very high,
experimentally unaccessible, energy scales. Concretely, since the
introduction of Right-handed (RH) neutrinos seem to be the obvious
addition needed in order to write a Dirac mass for the neutrinos,
and the seesaw can be used to explain the smallness of the neutrino
mass scale, most models assume their existence with a mass scale
typically of size $\sim 10^{13-16}$~GeV~\cite{seesaw2,seesaw3}.

In this paper we adhere to the idea that our current (experimental)
knowledge of particle physics should be explored by a "truly
minimal" extension of the Standard Model (SM). In this tenor we
consider the possibility of having only one scale associated with
all the high energy physics (HEP) phenomena. Since the SM is
consistent with all data so far (modulo neutrino masses), we propose
a minimal extension of the SM where new phenomena associated to
neutrino physics can also be explained by physics at the Electroweak
(EW) scale which we take to be in the range from $10$~GeV to $1$~TeV
(similar approaches can be found
in~\cite{hnus1,similar1,similar2,similar3}). Thus, we assume
\begin{itemize}
    \item SM particle content and gauge interactions.
    \item Existence of three RH neutrinos with a mass scale of
    EW size.
    \item Global U(1)$_L$ spontaneously (and/or explicitly) broken
    at the EW scale by a single complex scalar field.
    \item All mass scales come from spontaneous symmetry breaking
    (SSB). This leads to a Higgs sector that includes a Higgs
    SU(2)$_L$ doublet field $\Phi$ with hypercharge $1$
    (i.e. the usual SM Higgs doublet) and a SM singlet complex
    scalar field $\eta$ with lepton number $-2$.
\end{itemize}

This approach will have an effect on the type of signals usually
expected from the Higgs sector of the SM, where the hierarchy
(naturalness) problem resides. By enlarging the SM to explain the
neutrino experimental results, we can get a richer spectrum of
signals for Higgs physics and it is expected that once the LHC
starts, it will allow us to test some of the theoretical frameworks
created thus far. In any case, in order to fully probe whether the
Higgs bosons have ``Dirac'' and/or ``Majorana'' couplings, we might
have to wait until we reach a ``precision Higgs era'' at a linear
collider~\cite{lc}.

\section{The model}
Taking into account the previous assumptions it is straightforward
to write the Lagrangian. The relevant terms for Higgs and neutrino
physics are
\beq \label{lagrangian}
    {\cal L}_{\nu H}={\cal L}_{\nu y} - V \ ,
\eeq
with
\beq \label{yukawas}
    {\cal L}_{\nu y}= -y_{\alpha i}\bar{L}_{\alpha}N_{Ri}\Phi
    -\frac{1}{2}Z_{ij}\eta \bar{N}_{Ri}^cN_{Rj} + h.c. \ ,
\eeq
where $N_R$ represents the RH neutrinos, $\psi^c=C\gamma^0\psi^*$
and $\psi_R^c\equiv(\psi_R)^c=P_L\psi^c$ has left-handed chirality.
The potential is given by
\beq
\label{potential}\nonumber
    V & = &
    \mu_D^2\Phi^{\dagger}\Phi+\frac{\lambda}{2}\left(\Phi^{\dagger}\Phi\right)^2
    + \mu_S^2\eta^*\eta +
    \lambda^{\prime}\left(\eta^*\eta\right)^2  \\
    & + & \kappa \left(\eta\Phi^{\dagger}\Phi + h.c. \right)
    +\lambda_m \left(\Phi^{\dagger}\Phi\right)\left(\eta^*\eta\right) \ .
\eeq
Note that the fifth term in the potential breaks explicitly the
U(1) associated to lepton number. This is going to be relevant when
we consider the Majoron later in the paper.

Assuming that the scalar fields acquire vacuum expectation values
(vevs) in such a way that $\Phi$ and $\eta$ are responsible for EW
and global U(1)$_L$ symmetry breaking respectively, and using the
notation
\beq \label{vevs}
    \Phi = \left( \begin{array}{c} 0 \\ \frac{\phi^0 + v}{\sqrt{2}}
    \end{array} \right) \ \ {\rm and} \ \ \eta = \frac{\rho + u + i
    \sigma}{\sqrt{2}} \ ,
\eeq where $v/\sqrt{2}$ and $u/\sqrt{2}$ are the vevs of $\Phi$ and
$\eta$ respectively, we obtain the following minimization
conditions: \beq\label{minimization}
    \mu_D^2 & = & -\frac{1}{2} \left( \lambda v^2 + \lambda_m u^2
    -2\sqrt{2} \kappa u \right) \\
    \mu_S^2 & = & -\frac{1}{2u} \left( 2\lambda^{\prime} u^3 +
    \lambda_m u v^2 + \sqrt{2} \kappa v^2 \right) \ .
\eeq
We can also obtain the mass matrix for the scalar fields and it
is given by
\beq \label{scalarmass}
    M_S^2= \left(\begin{array}{cc}
    \lambda v^2 & v u(\lambda_m  - \sqrt{2} r) \\ v u(\lambda_m
    -\sqrt{2} r) & 2\lambda^{\prime}u^2+\frac{1
    }{\sqrt{2}}r v^2 \end{array}\right) \ ,
\eeq where $r\equiv -\kappa/u$. The mass for the $\sigma$ (Majoron)
field is \beq \label{sigmamass} M_{\sigma}^2=\frac{r v^2}{\sqrt{2}}
\ . \eeq Note that, as expected, $M_{\sigma}^2$ is proportional to
the parameter $\kappa$ associated to the explicit breaking of the
U(1)$_L$ symmetry.

We are working under the assumption that the explicit breaking is
very small, i.e. $\kappa <<$ EW scale. This is why we are minimizing
the potential with respect to $\eta$ thus assuming it does break the
symmetry spontaneously. Furthermore we expect the SSB generated by
the vev of $\langle \eta \rangle = u$ to be of EW scale size and so
we work under the assumption $r\equiv -\kappa/u << 1$. For example,
taking $-\kappa \sim$~KeV one obtains $r \sim 10^{-7 -9}$ which then
leads to a Majoron mass of hundreds of KeV.

From Eq.(\ref{scalarmass}) we see that it is useful to define the
mass eigenstates \beq \label{scalarfield}
     {\cal H} = \left( \begin{array}{c}
     \phi^0 \\ \rho \end{array} \right) =
     \left( \begin{array}{cc}
     \cos\alpha & -\sin\alpha \\ \sin\alpha & \cos\alpha \end{array} \right)
     \left( \begin{array}{c}
     h \\ H \end{array} \right) \ .
\eeq

Using these definitions to rewrite Eq.(\ref{yukawas}) we obtain \beq
\label{yukawas2} \nonumber
     {\cal L}_{\nu y} & \supset & -y_{\alpha i} \bar{\nu}_{L\alpha}N_{Ri}\frac{\phi^0}{\sqrt{2}}
     -\frac{1}{2} Z_{ij} \frac{(\rho+i\sigma)}{\sqrt{2}} \bar{N}_{Ri}^c N_{Rj} + h.c. \\ \nonumber
     & = & \left( -\frac{y_{\alpha i}}{\sqrt{2}}\bar{\nu}_{L\alpha}N_{Ri}
     (c_{\alpha} \ h -s_{\alpha} \ H) + h.c. \right) - \left(\frac{i}{2\sqrt{2}}
     Z_{ij} \bar{N}_{Ri}^c N_{Rj} \sigma + h.c. \right) \\
     & - & \left( \frac{1}{2\sqrt{2}} Z_{ij} \bar{N}_{Ri}^c N_{Rj} (s_{\alpha} \ h +
     c_{\alpha} \ H) + h.c \right) \ .
\eeq

We now make some comments regarding neutrino mass scales. Since we
are interested in RH neutrinos at the EW scale, we take their masses
to be in that scale, i.e. anywhere from a few to hundreds of GeV.
The Dirac part on the other hand will be constrained from the
seesaw. Writing the neutrino mass matrix as
\beq \label{mneutrino} m_{\nu} = \left( \begin{array}{cc} 0 & m_D \\
m_D & M_M \end{array}\right) \ , \eeq where $(m_D)_{\alpha i} =
y_{\alpha i}v/\sqrt{2}$.  As an example lets consider the third
family of SM fields and one RH neutrino, thus Eq.({\ref{mneutrino})
becomes a $2\times 2$ matrix. Assuming $m_D << M_M$ we obtain the
eigenvalues $m_1=-m_D^2/M_M$ and $m_2=M_M$ and by requiring $m_1
\sim$~O(eV) and $m_2 \sim (10 -100)$~GeV and using $v=246$~GeV we
obtain an upper bound estimate for the coupling $y_{\tau i} \leq
10^{-6}$.

The mass eigenstates are denoted by $\nu_1$ and $\nu_2$ and are such that
\beq \label{transformation} \nonumber
\nu_{\tau} & = & \cos\theta \ \nu_{L1} + \sin\theta \ \nu_{R2} \\
N & = & -\sin\theta \ \nu_{L1} +\cos\theta \ \nu_{R2} \ , \eeq where
$\theta=\sqrt{m_D/m_2} \approx 10^{-(5 - 6)}$.

The relevant terms in the Lagrangian become \beq
\label{relevantterms} \nonumber {\cal L} & \supset & \left[ h
\bar{\nu}_{L1}^c \nu_{L1} \left( -\frac{Z}{2\sqrt{2}}s_\theta^2
s_\alpha \right)+  h \bar{\nu}_{R2}^c \nu_{R2}
\left(- \frac{Z}{2\sqrt{2}} c_\theta^2 s_\alpha \right) + h.c. \right] \\
& + & h \bar{\nu}_{L1} \nu_{R2}
\left(\frac{y_\nu}{\sqrt{2}}(s_{\theta}^2-c_{\theta}^2)c_\alpha
\right) +  h \bar{\nu}_{R2} \nu_{L1}
\left(\frac{y_\nu}{\sqrt{2}}(s_\theta^2-c_{\theta}^2) c_\alpha
\right) \ , \eeq where $y_{\nu}^*=y_{\nu}$ and $Z \equiv Z_{11}$.

As discussed in the introduction we are interested in exploring the
Higgs decays to neutrinos and their signatures in this model. The
possible decay modes involving the Majoron and its relation to Dark
Matter will be considered elsewhere. Then, using
Eq.~(\ref{relevantterms}) we compute the following decay
widths~\footnote{All SM decay widths will include an extra factor of
$c_{\alpha}^2$}: \beq \label{widths}
\Gamma(h\rightarrow \bar{\nu}_1\nu_1) & = & \frac{m_h}{64\pi}|Z|^2 s_{\theta}^4 s_{\alpha}^2 \ , \\
\Gamma(h\rightarrow \bar{\nu}_2\nu_2) & = & \frac{m_h}{64\pi}|Z|^2 c_{\theta}^4 s_{\alpha}^2
\left(1-\frac{4m_2^2}{m_h^2}\right)^{3/2} \ , \\
\Gamma(h\rightarrow \bar{\nu}_1\nu_2) & = &
\frac{m_h}{16\pi}y_{\nu}^2(s_{\theta}^2-c_{\theta}^2)^2 c_{\alpha}^2
\left(1-\frac{m_2^2}{m_h^2}\right)^2 \ . \eeq

\section{Numerical results}
We have computed the branching ratios for the Higgs decays and the
results are presented in Figure~\ref{fig:branching}. In each plot we
have included the results for three values of $\cos\alpha$ ($0.1$,
$0.5$ and $0.9$). The three graphs correspond to the values of $m_2
= 10$, $60$ and $100$~GeV respectively. Only the dominant
contributions are shown for clarity, i.e. $h \to \nu_2\bar{\nu_2}, \
ZZ, \ WW, \ b\bar{b} \ \rm{and} \ \tau \bar{\tau}$. It is interesting to note
that for the whole range where it is possible, the decay $h \to
\nu_2 \bar{\nu_2}$ dominates in all three cases for small $\cos\alpha$ and
it is still relevant for large $\cos\alpha$. This is a clear
distinctive signature of our model.

In order to study the specific signatures that would be observed in
this scenario, we consider the $\nu_2$ decays. In
Table~\ref{tab:signatures} we present the possible signatures of
these decays.
\begin{figure}[ht]
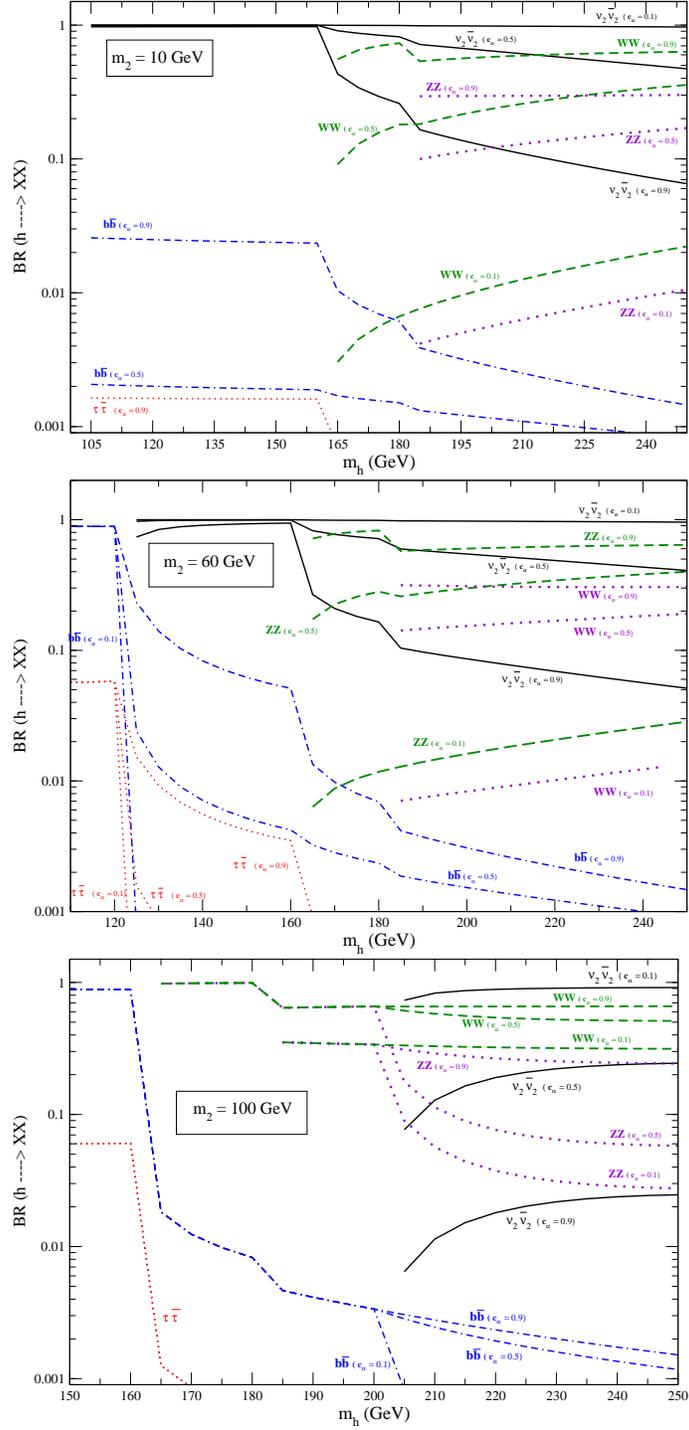

  \begin{center}
    \includegraphics[width=9cm]{fig1.eps}
    \includegraphics[width=9cm]{fig2.eps}
    \includegraphics[width=9cm]{fig3.eps}
    \caption{Dominant branching ratios for Higgs decays.
    Three cases are presented for $m_2 = 10, \ 60 \ \rm{and} \ 100$~GeV
    respectively. Each plot includes results for the three values of
    $\cos\theta = 0.1, \ 0.5 \ \rm{and} \ 0.9$ as discussed in the text.}
    \label{fig:branching}
  \end{center}
\end{figure}

\begin{table}[ht]
  \begin{center}
    \begin{tabular}{||c|c|c|c||}
      \hline
      {\rm Higgs decay} & $\nu_2 \rightarrow \nu_1 Z^*$ &
      $\nu_2 \rightarrow l W^*$ &  $\nu_2 \rightarrow \nu_1 \gamma$ \\
      \hline
      \hline
      $h\rightarrow \nu_1\nu_2$ & $l^+l^- + {\rm inv.}$ & $l+l^{\prime}+ {\rm inv.}$ &
      $\gamma + {\rm inv.}$ \\
      & $q\bar{q} + {\rm inv.}$ & $l+q\bar{q}^{\prime}  + {\rm inv.}$ & \\
      \hline
      $h\rightarrow \nu_2\nu_2$ & $l^+l^- + l^+l^- + {\rm inv.}$ &
      $l + l^{\prime} + l^{\prime\prime}+ l^{\prime\prime\prime} + {\rm inv.}$ & \\
      & $l^+l^- + q\bar{q} + {\rm inv.}$ & $l + l^{\prime} + l^{\prime\prime}+ q\bar{q}  + {\rm inv.}$ &
      $\gamma + \gamma  + {\rm inv.}$ \\
      & $q\bar{q} + q\bar{q}  + {\rm inv.}$ & $l+l^{\prime}+ q\bar{q}+q\bar{q} + {\rm inv.}$ & \\
      \hline
      $h\rightarrow \nu_1\nu_1$ & - & - & - \\
      \hline
    \end{tabular}
    \caption{Signatures for the Higgs decays considered in the text.}
    \label{tab:signatures}
  \end{center}
\end{table}

Since we are interested in a Higgs mass in the natural window of
$100-200$~GeV, and in neutrino masses such that they can appear in
Higgs decays, we will consider neutrino masses of order
$10-100$~GeV, therefore we need to consider the 2 and 3-body decays 
$\nu_2 \to V + l$ and $\nu_2 \to \nu_1 + V^* (\to f \bar{f}')$, 
where $V^*= W^* , Z^*$ 

One can also evaluate the branching ratios for the neutrino radiative decay,
but since this is a loop-process, it is quite suppressed unless the
mass differences among the right and left-handed neutrinos are very
small. We have not included such process in this work.

Consider the process in Figure~\ref{fig:Ndecay}. Its decay width is
given by 
\beq \label{widthN}
\Gamma=\frac{m_2^5}{384\pi^3M_v^4}\left[(B^2+C^2)(a_f^2+b_f^2)\right] \ , 
\eeq where 
\beq \label{parameters} \nonumber (V = W) &
\rightarrow & \left\{ \begin{array}{c}
a_f  =  -b_f \equiv a = \frac{g}{2\sqrt{2}} \\
B = -C = a \ s_{\theta} \end{array} \right. \\
(V = Z) & \rightarrow & \left\{
\begin{array}{c}
a_f = \frac{g}{2c_w}(T^3_f-2Q_f s_w^2) \\ \nonumber
b_f = -\frac{g}{2c_w} T^3_f \\
B = a_{\nu} \ c_{\theta}s_{\theta} \\
C = b_{\nu} \  c_{\theta}s_{\theta} \end{array} \right. \ . 
\eeq
 For the 2-body decays the result is
\beq\label{2body} 
\Gamma=\frac{(B^2+C^2)(m_2^2-M_v^2)^2(1+2\frac{M_v^2}{m_2^2})}
{8\pi M_v^2 m_2^2} \ .
\eeq

\begin{figure}[ht]
  \begin{center}
    \includegraphics[width=9cm]{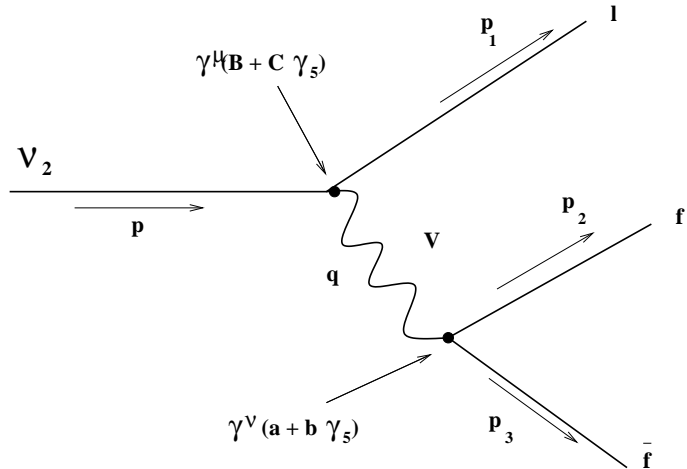}
    \caption{Three body decay for $\nu_2$.}
    \label{fig:Ndecay}
  \end{center}
\end{figure}

We have evaluated the branching ratios for these processes and they
are presented in table~\ref{tab:nu2decays}. We show the results for
$m_2 = 100$~GeV as the results are similar in all the $m_2$ range
considered in this paper. We find that the dominant contributions
are the ones associated to the $W^*$ decay process.
\begin{table}[ht]
    \begin{center}
        \begin{tabular}{||c||c|c||c|c|c|c|c|c||}
            \hline
            \hline
            $m_2 (GeV)$ & $Z \ \nu$ & $W \ l$ & $\nu \ l^+ \ l^-$ & $\nu \ \nu \ \nu$ & $\nu
            \ q_u \ \bar{q}_u$ & $\nu \ q_d \ \bar{q}_d$ & $l^{\pm} \ l^{\pm} \ \nu$ &
            $l^{\pm} \ q \ \bar{q}^{\prime}$ \\
            \hline
            \hline
            $10$ & - & - & $0.013$& $0.025$ & $0.029$ & $0.055$ & $0.293$ &
            $0.586$ \\
            $60$ & - & - & $0.013$& $0.025$ & $0.029$ & $0.055$ & $0.293$ &
            $0.586$ \\
            $100$ & $0.117$ & $0.883$ & $0.006$& $0.005$ & $0.126$ & $0.024$ & $0.294$ &
            $0.589$ \\
            \hline
        \end{tabular}
        \caption{Branching ratios for the $\nu_2$ two and three body decays discussed in the text.}
        \label{tab:nu2decays}
    \end{center}
\end{table}

\section{Discussion and conclusions}
A longstanding question in neutrino physics has been to determine
whether neutrino masses are of Dirac or Majorana type, which in turn
motivates the terminology used in calling the neutrinos either Dirac
or Majorana. As the Higgs mechanism employed in this paper is at the
root of the origin of both types of masses, we find it reasonable to
ask the same question for the Higgs couplings, namely, we would like
to determine wether the Higgs couplings are dominated by its Dirac
or Majorana components.

In fact, the interaction eigenstates that appear in our model,
$\Phi^0$ and $\eta^0$, do have well defined couplings to neutrinos:
of Dirac type the former and Majorana type the latter. Although one
may think that such a question is academic, we argue that it is not
the case, and that it will be possible to study the experimental
signatures that would distinguish among both types of couplings at
coming colliders.

At the base of our discussion is the fact that the Dirac couplings
$\phi^0 \bar{\nu}_l \nu_R$, involve both types of chiralities (L and
R), whereas the Majorana one $\eta^0 \nu^c_R \nu_R$, involves only
one chirality. Therefore, in a Higgs decay of Dirac type, one would
have a fermion of a given chirality and an anti-fermion with the
opposite chirality, while in the Majorana case, the Higgs decay
would involve a fermion pair of like chiralities. In our model, as
the decays $h\to \nu_1 \nu_1$ would escape detection, while the
decay $h\to \nu_1 \nu_2$ will only have one detectable neutrino,
which does not allow the possibility to correlate chiralities, we
are left only with the decay:  $h\to \nu_2 \nu_2$. Let us follow the
decay chain produced after the neutrino decays into a lepton and a
pair of jets, namely $\nu_2 \to l+ qq'$. It is then possible to a
have a pair of same-sign charged leptons, plus jets, which should
help in order to discriminate against backgrounds. Furthermore, the
charged leptons will inherit the chiralities of the neutrinos, and
its measurement will allow to test the Higgs couplings. A detailed
simulation study is needed, but this is beyond the scope of the
present paper.

We close this discussion with a few comments on the possibility to
measure the Dirac or Majorana coupling of the Higgs bosons in models
with a richer Higgs spectrum. For instance we could have models with
Higgs triplets that will include double-charged Higgs states
$\delta^{++}$, which then couple to a lepton pair, therefore
violating lepton number. As such state would decay into $e^+e^+$ or
$\mu^+ \mu^+$, it would then be possible to measure the chiralities
of those light leptons appearing in the final state, and therefore
to test the Dirac or Majorana nature of their couplings.

It is quite interesting that the Higgs sector presented in this
paper can lead to substantial modification of the signatures of the
Higgs bosons. Although these signatures seem quite different from
those expected in the SM, they represent the kind of variations
that one would have when new physics beyond the SM is included,
which in our case is well motivated by the plethora of recent
neutrino physics experiments. As the LHC is expected to start
operation very soon, it is very important to have an open mind
regarding possible variations of the signals expected from the SM
Higgs.

\begin{acknowledgments}
A.A. and J.L.D.C acknowledge support from CONACYT and SNI (Mexico).
\end{acknowledgments}

\end{document}